\documentstyle[amssymb,11pt]{article}
\input{psfig}
\begin{document} 
\begin{flushright}
{OUTP-00-77-P}\\ 
\end{flushright}   
\vskip2 cm

\begin{center} 
{\Large {\bf Precision prediction of gauge couplings and the profile of a 
string theory. }\\[0pt] 
\bigskip } \vspace{0.73cm} {\large {\bf Dumitru M. Ghilencea$^{a,b}$ and 
Graham G. Ross$^{a}$} \bigskip }\\[0pt] 
$^{a}${\it Department of Physics, Theoretical Physics, University of Oxford}%
\\[0pt] 
{\it 1 Keble Road, Oxford OX1 3NP, United Kingdom}\\[0pt] 
\vspace{0.23cm} $^{b}${\it Physikalisches Institut der Universitat Bonn,} \\[%
0pt] 
{\it Nussallee 12, 53115 Bonn, Germany}\\[0pt] 
 
\bigskip \vspace{1cm} {\bf Abstract} 
\end{center}

{\small We estimate the significance of the prediction for the gauge 
couplings in the MSSM with an underlying unification. The correlation 
between the couplings covers only (0.2-2)\% of the a priori 
reasonable region of the parameter space, 
while the prediction for $\sin ^{2}\theta _{W}$  is accurate to 
1.3\%. Given that agreement with experiment to such precision is unlikely to 
be fortuitous, we discuss the profile of a string theory capable of 
preserving this level of accuracy. We argue that models with a low scale of 
unification involving power law running in the gauge couplings do not. Even 
theories with a high scale of unification are strongly constrained, 
requiring the compactification scale of new space dimensions in which states 
transforming under the Standard Model propagate to be very close to the 
string cut-off scale. As a result {\it no} new space dimensions }can be 
larger than $10^{-14}fm.$\newpage  
 
\section{Introduction} 
 
The unification of gauge couplings is one of the few believable precision 
predictions coming from physics beyond the Standard Model. In the Minimal 
Supersymmetric Standard Model the radiative corrections to the gauge 
couplings coming from the states of the MSSM causes the SU(3), SU(2) and 
U(1) gauge couplings to become very nearly equal at the scale $(1-3).10^{16}$ 
GeV, provided one adopts the SU(5) normalisation of the U(1) factor. It is 
usual to use gauge unification to predict the value of the strong coupling 
given the weak and electromagnetic couplings. Doing this one finds 
$\alpha_{3}(M_{Z})=0.126\pm 0.01$ to be compared with the experimental value 
\cite{particledata} $\alpha _{3}(M_{Z})=0.118\pm 0.002$. However, the gauge 
unification prediction is really much more precise than this comparison 
suggests. In the MSSM the precision is limited by threshold effects 
associated with the masses of the supersymmetry partners of Standard Model 
states affecting the prediction for $\alpha _{3}(M_{Z})$. In Section \ref 
{precision} we discuss these corrections and demonstrate that, expressed as 
a correlation between the gauge couplings, the prediction covers only 
(0.2-2)\% of the overall (perturbative) 
parameter space. Moreover the prediction for  $\sin^{2}\theta_{W}$ 
is much more precise than for $\alpha _{3}$, being accurate 
to $1.3\%$. The fact that experiment and theory agree to this precision seems 
to us unlikely to be just a happy accident. It is largely due to this 
circumstantial evidence for unification and the fact that supersymmetry is 
needed to solve the mass hierarchy problem that there has been so much 
interest in supersymmetric models and in compactification schemes that 
preserve a stage of low energy supersymmetry. For the same reason it is of 
interest to demand that the physics beyond the MSSM should not spoil the 
prediction. 
 
There are inevitably further threshold corrections coming from new states 
lying at and above the unification scale. In the case of Grand Unification 
the states include the heavy gauge bosons and Higgs bosons needed to make up 
complete GUT multiplets. The threshold corrections associated with these 
states have been extensively discussed in the context of Grand Unification  
\cite{daniels} and the condition that the gauge unification relations should 
not be significantly affected requires there should be no substantial 
splitting between the heavy Higgs and gauge boson states. Provided the ratio 
between these masses is less than a factor of 10 the precision is 
essentially unchanged in minimal GUT schemes, although the requirement of 
near equality of Higgs and gauge boson masses becomes much stronger for the 
case of non-minimal Higgs sectors responsible for breaking the GUT. While at 
one loop order the unification scale is well defined in a SUSY GUT, at two 
loop order the scale of unification is not so easy to define. We discuss 
this point in Section \ref{twoloop} and show that it is necessary to 
regulate the theory to determine these two loop corrections. In a string 
theory this regularisation introduces new threshold corrections (associated 
with the canonical normalisation of the K\"{a}hler term for matter fields%
\footnote{%
Additional corrections due to the rescaling of the metric (super-Weyl 
anomaly) exist, with similar implications \cite{dixon2}.}) and a sensitivity 
to the string (cut-off) scale. As a result there are new contributions to 
the gauge couplings in Grand Unified theories proportional to $%
ln(M_{s}/M_{GUT})$ with the string scale, 
$M_{s}$ providing the cut-off scale, which are not 
usually included in GUT calculations. 
 
In the case of compactified string theories (with or without Grand 
Unification below the string scale) the threshold corrections can be much 
larger if the compactification scale is (significantly) different from the 
string scale. This may be understood (in the effective low energy field 
theory) as due to Kaluza Klein states associated with gauge non singlets 
propagating in the extra dimensions and with mass up to the string scale  
\cite{string} which contribute to the running of the gauge couplings. For 
low compactification scales the number of such states is very large leading 
to very large threshold corrections generating power-like gauge coupling 
running and significant threshold sensitivity of the predictions \cite 
{extradimensions}. String calculations of the threshold effects for the 
weakly coupled heterotic string have been performed by several groups (for a 
review of the topic and references see \cite{nilles1}). The regularisation 
of the effective field theory is provided at the (heterotic) string level by 
the geometry of the string\footnote{%
i.e. the world-sheet torus, \cite{string}}. In type I models with compact 
``in-brane'' dimensions \cite{kdienes} one may have a similar (power-like) 
behaviour due to the associated Kaluza Klein states. Finally, power-like 
behaviour may also be present in type I models with {\it anisotropic} 
compactification. The string thresholds have been computed in \cite{dudas} 
and (string) regularisation is again provided by the geometry of the string%
\footnote{%
via cancellation of the quadratic terms between the annulus and Moebius 
strip contributions, \cite{dudas}.}. As we shall discuss in Section 3 the 
expectation is that all these corrections will spoil the MSSM prediction and 
imply the compactification scale(s) should be very close or equal to the 
string scale. One important implication of this is that one cannot appeal to 
non-standard ``power law'' running to lower the unification scale and one is 
left only with high scale unification. 
 
The need for a high unification (and string) scale is broadly in agreement 
with the expectation in the heterotic string. However in the weakly coupled 
case there is a residual discrepancy of approximately a factor of 20 between 
the gauge unification scale found in MSSM unification and the string 
prediction. Further the requirement that the compactification scale should 
also be close to the string scale apparently prevents large threshold 
effects from coming to the rescue. However, as we discuss in Section \ref 
{wilson}, in cases of string thresholds with a Wilson line background, it is 
possible for large Wilson line effects to lower the unification scale to 
that of the MSSM while keeping the compactification scale close to the 
string scale. Thus the weakly coupled heterotic string with Wilson line 
background provides an example of a string model which can preserve the MSSM 
precision prediction and have an acceptable scale of unification. Another 
possibility is that the string theory has a stage of Grand Unification below 
the compactification scale allowing for a gauge unification scale below the 
compactification scale. As we discuss in Section \ref{twoloop}, even in this 
case, it is necessary that the compactification scale should be close to the 
string cut-off scale due to the need to keep two-loop corrections under 
control. An alternative resolution of the discrepancy is that gravity can 
propagate in more dimensions than matter and gauge states, changing the 
string prediction for the unification scale. The original idea for this came 
from the strongly coupled heterotic string in which the gauge unification 
scale is lowered through the appearance of an 11th dimension \cite{witten}. 
Even in this case the compactification scale of the other compactified 
dimensions cannot be far from the string scale. Our conclusions are 
presented in Section \ref{profile} where we present a profile of a 
compactified string theory capable of satisfying all the constraints needed 
to preserve the accuracy of the predictions for gauge couplings. 
 
\vspace{0.2cm} 
 
\section{Precision gauge unification in the MSSM\label{precision}} 
 
In the MSSM the prediction of the gauge couplings follows from the GUT or 
string relation at the unification scale corrected by the renormalisation 
group determination of radiative corrections involving the MSSM states. By 
combining the standard model gauge couplings one may eliminate the one-loop 
dependence on the unification scale and the value of the unified coupling to 
obtain a relation which depends only on the threshold effects associated 
with the unknown masses of the supersymmetric states.  
\begin{equation} 
\alpha _{3}^{-1}(M_{Z})=\left\{ \frac{15}{7}\sin ^{2}\theta _{W}(M_{Z})-%
\frac{3}{7}\right\} \alpha _{em}^{-1}(M_{Z})+\frac{1}{2\pi }\frac{19}{14}\ln  
\frac{T_{eff}}{M_{Z}}+(two-loop)  \label{mssm} 
\end{equation} 
Note that the leading dependence on the SUSY thresholds comes from $T_{eff}$%
, an effective supersymmetric scale, while two loop corrections, which have 
a milder dependence on the scale, through (gauge and matter) wavefunction 
renormalisation $\ln Z\sim \ln \alpha _{i}(scale)$ can be ignored to a first 
approximation. In terms of the individual SUSY masses $T_{eff}$ is given by  
\cite{carena}  
\begin{equation} 
T_{eff}=m_{\widetilde{H}}\left( \frac{m_{\widetilde{W}}}{m_{\widetilde{g}}}%
\right) ^{\frac{28}{19}}\left( \frac{m_{H}}{m_{\widetilde{H}}}\right) ^{%
\frac{3}{19}}\left( \frac{m_{\widetilde{W}}}{m_{\widetilde{H}}}\right) ^{%
\frac{4}{19}}\left( \frac{m_{\widetilde{l}}}{m_{\widetilde{q}}}\right) ^{%
\frac{9}{19}}  \label{Teff} 
\end{equation} 
provided the particles have mass above $M_{Z}.$ The precision of the 
relation, eq.(\ref{mssm}), between Standard Model couplings is limited by 
the dependence on $T_{eff}$. However the uncertainty in $T_{eff}$ is limited 
by the fact that the supersymmetry masses are bounded from below because no 
supersymmetric states have been observed and from above by the requirement 
that supersymmetry solve the hierarchy problem. From eq.(\ref{mssm}) we see 
the dependence on the squark, slepton and heavy Higgs masses is very small, 
the main sensitivity being to the Higgsino, Wino and gluino masses. 

If all the supersymmetric partner masses are of O(1TeV) then so is $T_{eff}$.
However in most schemes of supersymmetry breaking radiative corrections
split the superpartners. In the case of gravity mediated supersymmetry
breaking with the assumption of universal scalar and gaugino masses at the
Planck scale the relations between the masses imply 
 $T_{eff}\simeq m_{%
\widetilde{H}}\left( \alpha _{2}(M_{Z}\right) /\alpha _{3}(M_{Z}))^{2}\simeq 
|\mu |/12,$.
For $\mu$ of order the weak scale $T_{eff}$ is approximately 
20 GeV. From this we
see the uncertainty in the SUSY breaking mechanism corresponds to a wide
range in $T_{eff}$. In what follows we shall take $20GeV<T_{eff}<1TeV$ 
as a reasonable estimate of this uncertainty.
Using this one may determine the 
uncertainty in the strong coupling for given values of the weak and 
electromagnetic couplings using eq.(\ref{mssm}). The result (including 
two-loop effects) as a function of $\alpha _{3}$ and $\sin ^{2}\theta _{W}$ 
is plotted in Figures 1(a) and 1(b)$\footnote{%
Further constraints on this region, coming from small fine-tuning of other 
observables of the MSSM \cite{bastero} may further constrain the area of 
allowed values of $\alpha _{3}(M_{Z})$ vs. $\sin ^{2}\theta _{W}$.}$.  
\begin{figure}[t] 
\label{mssmfigure0}  
\begin{tabular}{cc|cr|} 
\parbox{8cm}{ 
\psfig{figure=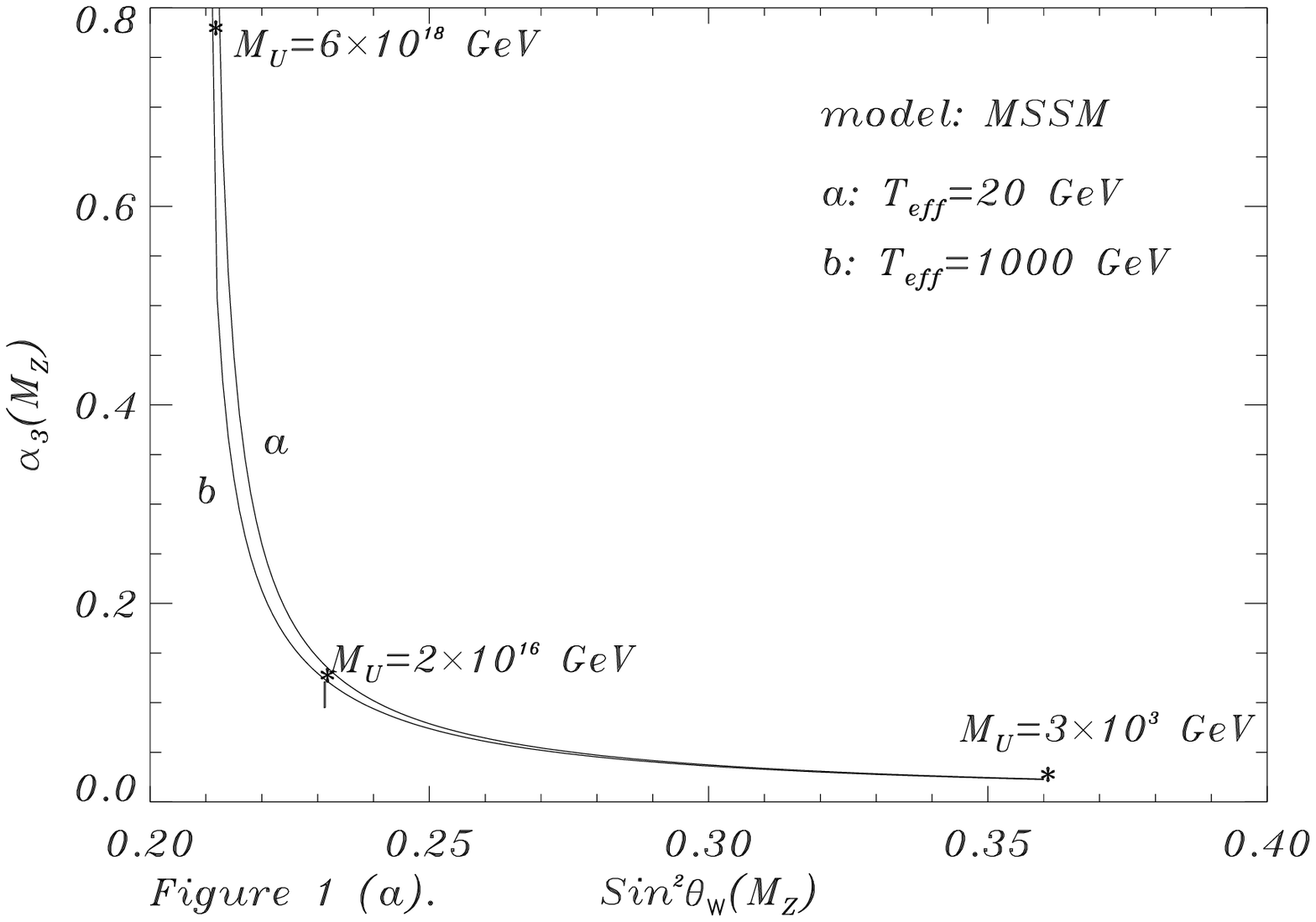,height=7.5cm,width=7.5cm}} 
\hfill{\,\,\,\,\,\,\,\,\,\,\,\,\,\,\,} 
\parbox{8cm}{ 
\psfig{figure=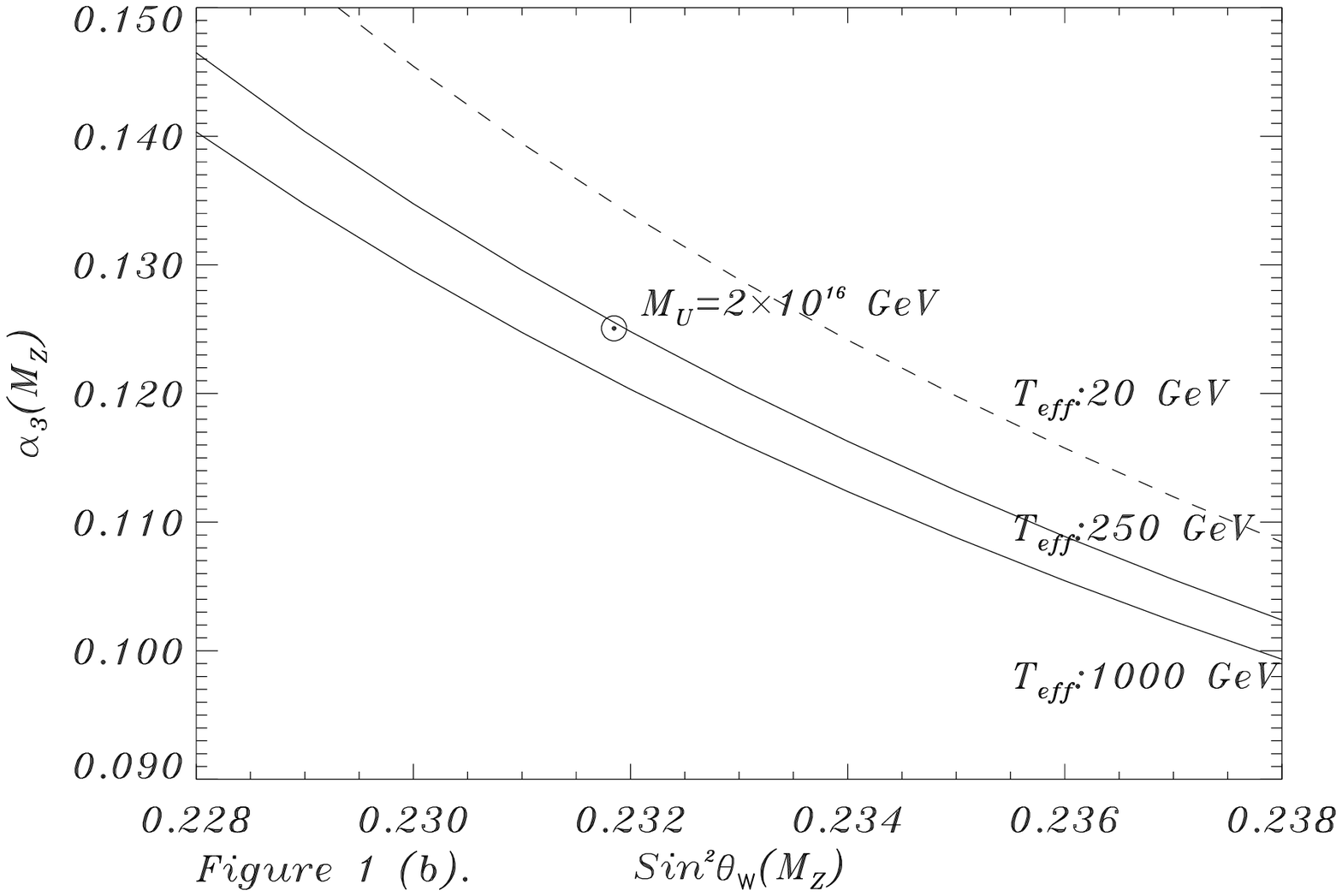,height=7.5cm,width=7.5cm}}  
\end{tabular} 
\newline 
\newline 
\newline 
{\small Figures 1 (a) and (b). {Plots of $\alpha _{3}(M_{Z})$ versus }}$\sin 
^{2}\theta _{W}$ calculated {\small {in the MSSM for two values of the 
effective supersymmetric threshold, $T_{eff}=20$ GeV and $T_{eff}=1000$ GeV. 
The limits correspond to requiring $\alpha _{3}(M_{Z})$ remains in the 
perturbative domain and unification occurring above the supersymmetry 
threshold. The area between the two curves provides a measure of the 
predictivity of the theory. The experimental range of values is also shown.}} 
\end{figure} 
 
The precision of the prediction is remarkable. A measure of this is given by 
the area between the two curves in Figure 1 (a). If one assumes that a 
random model not constrained by unification may give any value for $\alpha 
_{3}$ and $\sin ^{2}\theta _{W}$ between 0 and 1 the relative precision is 
an impressive 0.002! Of course this estimate is sensitive to the measure 
chosen. Changing to $\alpha _{3}$ and $\sin \theta _{W}$ makes very little 
difference. Changing to $\alpha _{3}^{-1}$ and $\sin ^{2}\theta _{W}$ (and 
restricting the possible range of $\alpha _{3}^{-1}$ to be $1<\alpha 
_{3}^{-1}<10$) increases the relative precision by a factor of 10. Using 
this variation as an estimate of the uncertainty associated with the measure 
we conclude that a reasonable estimate for the precision of the prediction 
for the correlation between the Standard Model couplings is in the range $%
(0.2-2)\%.$ One may also use the result of eq(\ref{Teff}) to predict one of 
the couplings given the other. However, as may be seen from Figure 1(a), the 
prediction is not equally precise for each coupling. A quantitative estimate 
of the accuracy of the prediction for $\alpha _{3}(M_{Z})$ or $\sin 
^{2}\theta _{W}(M_{Z})$ may be obtained from eq.(\ref{mssm})by taking the 
derivative with respect to $T_{eff}$ giving  
\begin{equation} 
{\alpha _{3}^{-1}(M_{Z})}\,{\cal F}_{\alpha _{3}}+\frac{15}{7}{\sin 
^{2}\theta _{W}(M_{Z})}\,{\alpha _{em}^{-1}(M_{Z})}\,{\cal F}_{\theta 
_{W}}\geq \frac{19}{28\pi }+two-loop  \label{bound} 
\end{equation} 
where  
\begin{equation} 
{\cal F}_{\alpha _{3}}=\left| \frac{d\ln (\alpha _{3}(M_{Z})}{d\ln T_{eff}}%
\right| ;\,\,\,\,\,\,\,{\cal F}_{\theta _{W}}=\left| \frac{d\ln (\sin 
^{2}\theta _{W})}{d\ln T_{eff}}\right|   \label{ratio} 
\end{equation} 
The quantities ${\cal F}_{\alpha _{3}}$ and ${\cal F}_{\theta _{W}}$ give 
the fractional threshold sensitivity of $\alpha _{3}$ and $\sin ^{2}\theta 
_{W}$ to a change in ln$T_{eff}$. From this we see that the unification 
prediction for $\sin ^{2}\theta _{W}(M_{Z})$ is more accurate than that for $%
\alpha _{3}(M_{Z})$ since, close to the experimental point, we have from 
eqs.(\ref{bound}) and (\ref{ratio})  
\begin{equation} 
\frac{{\cal F}_{\theta _{W}}}{{\cal F}_{\alpha _{3}}}\approx \frac{1}{7.6} 
\label{correlation} 
\end{equation} 
Given that the error in $\alpha _{3}$ is $\approx 
10 \%$ for a change in $T_{eff}$ 
from $20$ to $1000GeV$ we see  from this equation that the corresponding 
error in $\sin ^{2}\theta _{W}$ is $1.3 \%.$ Thus the prediction for 
$\sin^{2}\theta _{W}$ provides a more realistic measure of the
precision of  the unification prediction.
An expansion of $\sin^{2}\theta_{W}$ (corresponding to
$\alpha_3(M_z)=0.119$) thus gives 
\begin{equation} 
\sin ^{2}(\theta _{W}(M_{Z}))\simeq 0.2337-0.25(\alpha _{3}(M_{Z})-0.119)\pm 
0.0015 \label{sintheta} 
\end{equation} 
This effect may be seen directly 
in Figure 1(a) since the curve is more steeply 
varying in the $\alpha _{3}$ direction than in the $\sin^{2}\theta _{W}$ 
direction close to the experimental point. The optimal combination of 
$\alpha _{3}$ and $\sin^{2}\theta _{W}$ with minimal uncertainty  normal to 
the curve can be determined numerically but is relatively close to 
$\sin^{2}\theta _{W}.$ The 
fact that experiment and theory agree to this level of 
accuracy is impressive and is a major reason why so much attention has been 
paid to supersymmetric extensions of the Standard Model. 
 
\vspace{0.2cm} 
 
\section{High-scale threshold sensitivity of the unification prediction} 
 
Given this impressive accuracy it is clearly of interest to determine the 
effect on this prediction coming from threshold effects at the unification 
scale in realistic string theories. Our hope is that the requirement that 
the precision should not be significantly degraded will give us information 
about the underlying string theory. The value of the gauge couplings at $%
M_{Z}$ is of the form  
\begin{eqnarray} 
\alpha _{i}^{-1}(M_{Z}) &=&-\delta _{i}+\alpha ^{-1}(\Lambda )+\frac{%
\overline{b}_{i}}{2\pi }\Delta (\Lambda ,\mu _{0})+\frac{b_{i}}{2\pi }\ln  
\frac{\Lambda }{M_{Z}}  \nonumber \\ 
&&+\frac{3T_{i}(G)}{2\pi }\ln \left[ \frac{\alpha (\Lambda )}{\alpha 
_{i}(M_{Z})}\right] ^{1/3}-\sum_{\phi }\frac{T_{i}(R_{\phi })}{2\pi }\ln \ 
Z_{\phi }(\Lambda ,M_{Z})  \label{rge_general} 
\end{eqnarray} 
\linebreak where $\Delta $ is the string threshold correction corresponding 
to the particular string theory considered (hereafter renamed to $\Delta ^{H} 
$, $\Delta ^{I}$ to stand for the weakly coupled heterotic and type I string 
cases respectively). The parameter $\Lambda $ is the unification scale, $%
\alpha (\Lambda )$ is the unified coupling, $\mu _{0}$ is the mass of the 
heavy states, often the compactification scale in string theories and $%
Z_{\phi }\,$and ($\alpha (\Lambda )/\alpha (M_{Z})$) are the matter and 
gauge wavefunction renormalisation coefficients respectively. In eq(\ref 
{rge_general}) there are additional effects due to low energy supersymmetric 
thresholds $\delta _{i}$ and\footnote{%
The definition of $T_{eff}$ eq.(\ref{Teff}) in terms of $\delta _{i}$ is $%
\ln {T_{eff}}/{M_{z}}=-(28\pi /19)(\delta _{1}b_{23}/b_{12}+\delta 
_{2}b_{31}/b_{12}+\delta _{3})$, with $b_{ij}=b_{i}-b_{j}$.} $%
b_{i}=-3T_{i}(G)+\sum_{\phi }T_{i}(R_{\phi })$ are the one-loop beta 
function coefficients, while $\overline{b}_{i}$ depend on the particular 
compactification scheme - they are non-zero only for the N=2 massive SUSY 
states and vanish for the N=4 spectrum. 
 
In comparing the threshold effects at the unification scale to the SUSY 
threshold effects discussed above we will restrict ourselves to a measure of 
the sensitivity of the prediction to these scales for either the strong 
coupling or $\sin ^{2}\theta _{W}$ while the other is maintained fixed. In 
leading order we have from eq(\ref{rge_general})  
\begin{equation} 
\delta \alpha _{3}^{-1}(M_{Z})-\frac{15}{7}\delta \sin ^{2}\theta _{W}=\frac{%
1}{2\pi }\left[ \overline{b}_{1}\frac{b_{23}}{b_{12}}+\overline{b}_{2}\frac{%
b_{31}}{b_{12}}+\overline{b}_{3}\right] \delta \Delta (\Lambda ,\mu _{0}) 
\label{deltaalpha} 
\end{equation} 
where $b_{ij}=b_{i}-b_{j}$, $b_{1}=33/5$, $b_{2}=1$, $b_{3}=-3$. The 
relative sensitivity of $\alpha _{3}$ (keeping $\sin ^{2}\theta _{W}$ fixed) 
to changes in $\mu _{0}$ and $T_{eff}$ is, from eqs($\ref{sensitiv})$, (\ref 
{ratio}) and (\ref{bound}), given by  
\begin{eqnarray} 
{\cal {R}} &\equiv &\left| \frac{\delta (\ln (\alpha _{3}(M_{Z})))}{\delta 
(\ln (\alpha _{3}(M_{Z})))_{MSSM}}\right| =\left| \frac{d\ln \alpha 
_{3}(M_{Z})}{d\ln \mu _{0}}\right| \times {\cal {F}}_{\alpha _{3}}^{-1} 
\label{sensitiv} \\ 
&=&\frac{14}{19}\left| \left\{ \frac{5}{7}\overline{b}_{1}-\frac{12}{7}%
\overline{b}_{2}+\overline{b}_{3}\right\} \frac{d\Delta }{d(\ln \mu _{0})}%
\right| 
\end{eqnarray} 
where we have assumed that the predicted value for $\alpha _{3}(M_{Z})$ in 
the model considered is close to that of the MSSM\footnote{%
For a model to be viable one must in first instance predict the right value 
for $\alpha _{3}(M_{Z})$ and only after would the question of threshold 
sensitivity be relevant.}. One obtains the same result for the {\it relative} 
threshold sensitivity if we compute the prediction for $\sin ^{2}\theta 
_{W}(M_{Z})$ (with $\alpha _{3}(M_{Z})$ fixed) normalised to ${\cal F}%
_{\theta _{W}}$. For this reason in our analysis of threshold sensitivity we 
will focus our attention only on the value of ${\cal R}$. 
 
It follows immediately from eq(\ref{sensitiv}) that models involving only 
complete additional GUT representations (i.e. with $\overline{b}%
_{i}=b_{i}+n) $ do not introduce any threshold sensitivity at one loop 
order. This will be the case if there is a Grand Unification below the 
compactification scale. It is also true if one can find models for which $%
\overline{b}_{i}~=\kappa ~b_{i}$ as has been suggested \cite{dienes} to 
lower the unification scale by power law running. In both these cases, 
however, there will be significant threshold dependence at two loop order 
and we will discuss this separately in Section \ref{twoloop}. 
 
Returning to the one-loop threshold effects of eq.(\ref{sensitiv}), we will 
determine the sensitivity for the case of weakly coupled heterotic string 
with/without Wilson lines and type I/I$^{\prime }$ models to analyse their 
relative threshold sensitivity, ${\cal R}$. In what follows $\Lambda $, $%
\mu_{o}$ stand for the string and compactification scale respectively. To 
perform this analysis we use the explicit string computations for the string 
thresholds, $\Delta $ \cite{dixon,dudas,stieberger}. 
 
\subsection{Weakly coupled heterotic models} 
 
We first consider the case of the weakly coupled heterotic string with an 
N=2 sector (without Wilson lines present) in which six of the dimensions are 
compactified on an orbifold, $T^{6}/G$. In such models the spectrum splits 
into N=1, N=2 and N=4 sectors, the latter two associated with a $T^{2}\times 
T^{4}$ split of the $T^{6}$ torus. Due to the supersymmetric 
non-renormalisation theorem, the N=4 sector does not contribute to the 
running of the holomorphic couplings. As we shall discuss in Section 4 even 
in the case of N=4, two-loop corrections to the effective gauge couplings 
may introduce substantial threshold corrections. The N=1 sector gives the 
usual running associated with light states but does not contain any moduli 
dependence. The latter comes entirely from the N=2 sector. For the heterotic 
string all states are closed string states and at one loop the string world 
sheet has the topology of the torus $T^{2}$. For the case of a 
six-dimensional supersymmetric string vacuum compactified on a two torus $%
T^{2}$ the string corrections take the form \cite{kap,dixon}. Here $%
T\varpropto R_{1}R_{2}$ and $U\varpropto R_{1}/R_{2}$ where $T$, $U$ are 
moduli and $R_{1}$, $R_{2}$ are the radii associated with $T^{2}.$ We 
consider the case of a two torus $T^{2}$ with $T=iT_{2}$ (the subscript 2 
denotes the imaginary part) and $U=iU_{2}$. Making the dimensions explicit $%
T_{2}$ should be replaced by $T_{2}\rightarrow T_{2}^{o}/(2\alpha ^{\prime 
})\equiv 2R^{2}/(2\alpha ^{\prime })$. Performing a summation over momentum 
and winding modes and integrating over the fundamental domain of the torus 
gives the following result for $\Delta ^{H}$ \cite{dixon}  
\begin{eqnarray} 
\Delta ^{H} &=&-\frac{1}{2}\ln \left\{ \frac{8\pi e^{1-\gamma _{E}}}{3\sqrt{3%
}}\,U_{2}T_{2}{|}\eta (iU_{2}){|}^{4}\,\left| \eta \left( iT_{2}\right) 
\right| ^{4}\right\}   \label{threshold} \\ 
&=&-\frac{1}{2}\ln \left\{ 4\pi ^{2}\,{|}\eta (i){|}^{4}\left( \frac{M_{s}}{%
\mu _{0}}\right) ^{2}\left| \eta \left[ \frac{i\,3\sqrt{3}\pi }{2e^{1-\gamma 
}}\,\left( \frac{M_{s}}{\mu _{0}}\right) ^{2}\right] \right| ^{4}\right\}  
\label{di} 
\end{eqnarray} 
with the choice $U_{2}=1$ in the last equation. $M_{s}$ is the string scale,  
$\mu _{o}\equiv 1/R$, $\eta (x)$ is the Dedekind eta function and we have 
replaced $\alpha ^{\prime }$ in terms\footnote{%
>From \cite{kap} $M_{s}={2\,e^{(1-\gamma _{E})/2}\,3^{-3/4}}/\sqrt{2\pi 
\alpha ^{\prime }}
\label{mstring}$ where $g_{s}$ is the string coupling at the unification.} 
of $M_{s}$. For large\footnote{%
Note that $T_{2}\approx 5.5(M_{s}R)^{2}$ in ${\overline{DR}}$ scheme, so one 
can easily have $T_{2}\approx 20$ while $R$ is still close to the string 
length scale, to preserve the weakly coupled regime of the heterotic string. 
In this section ``large'' $R$ corresponds to values of $T_{2}$ in the above 
range.} $T_{2}$ the eta function is dominated by the leading exponential  
and so one finds the power law behaviour $\Delta \propto T_{2}\propto R^{2}$ 
which has a straightforward interpretation as being due to the 
decompactification associated with $T^{2}.$ This contribution is basically 
due to the tower of Kaluza Klein states below the string scale and can be 
understood at the effective field theory level \cite{string} whose result is 
regularised by the string world-sheet \cite{string}. The presence of 
power-like running, although taking place over a very small region of energy 
range\footnote{%
This takes place essentially between the compactification and the string 
scale \cite{string}}, can significantly affect the sensitivity of the 
unification prediction for $\alpha _{3}(M_{Z})$ with respect to changes of 
the compactification scale (which, being determined by a moduli vev, is not 
fixed perturbatively and hence is not presently known). To demonstrate this 
explicitly, consider the variation of the threshold $\Delta ^{H}$ with 
respect to $T_{2}$  
\begin{equation} 
\frac{\delta \Delta ^{H}}{\delta \ln T_{2}}=-\frac{1}{2}\left\{ 1+4\frac{%
d\ln |\eta (iT_{2})|}{d(\ln T_{2})}\right\}   \label{hetderiv} 
\end{equation} 
This gives the following relative threshold sensitivity (eq.(\ref{sensitiv}%
)) with respect to the compactification scale $\mu _{0}$ ($x\equiv \mu 
_{0}/M_{s}$)  
\begin{equation} 
{\cal {R}}=\frac{14}{19}\left| \left\{ \frac{5}{7}\overline{b}_{1}-\frac{12}{%
7}\overline{b}_{2}+\overline{b}_{3}\right\} \left\{ 1-2x\frac{d}{dx}\ln \eta 
\left( i\sigma x^{-2}\right) \right\} \right|  
\end{equation} 
where $\sigma =3\sqrt{3}\pi e^{\gamma _{E}-1}/2$. In Figure 2 we plot the 
second factor in curly brackets (which is equal to $\delta \Delta 
^{H}/\delta (\ln x)$)\footnote{%
This situation is somewhat similar to other (effective) models \cite 
{extradimensions} where it has been shown that power-like running brings in 
a significant amount of threshold sensitivity.}.  
\begin{figure}[t] 
\label{delta}  
\begin{tabular}{cc|cr|} 
\parbox{8cm}{ 
\psfig{figure=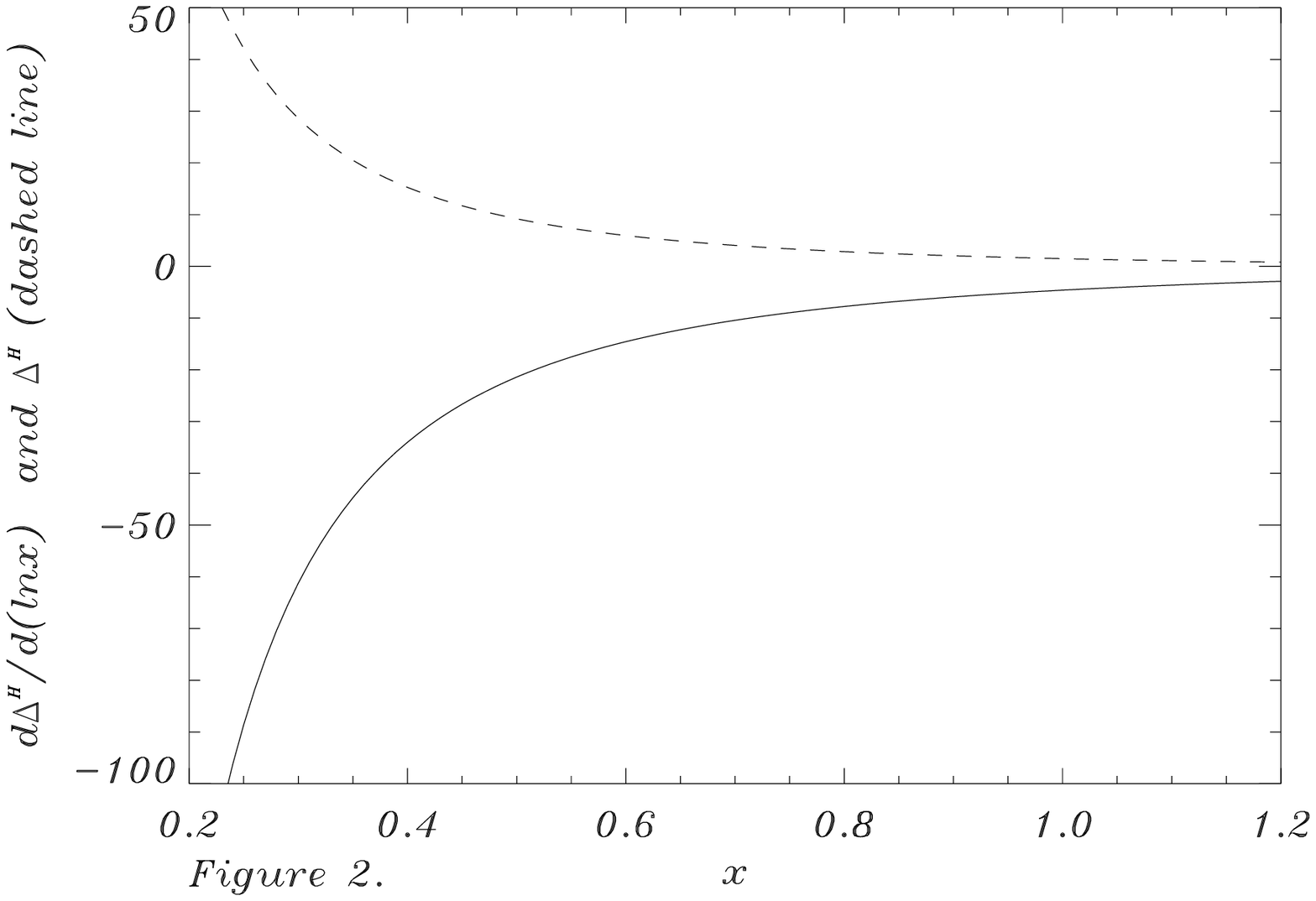,height=7.5cm,width=7.5cm}}  
\hfill{\,\,\,\,\,\,\,\,\,\,\,\,\,\,\,} 
\parbox{8cm}{ 
\psfig{figure=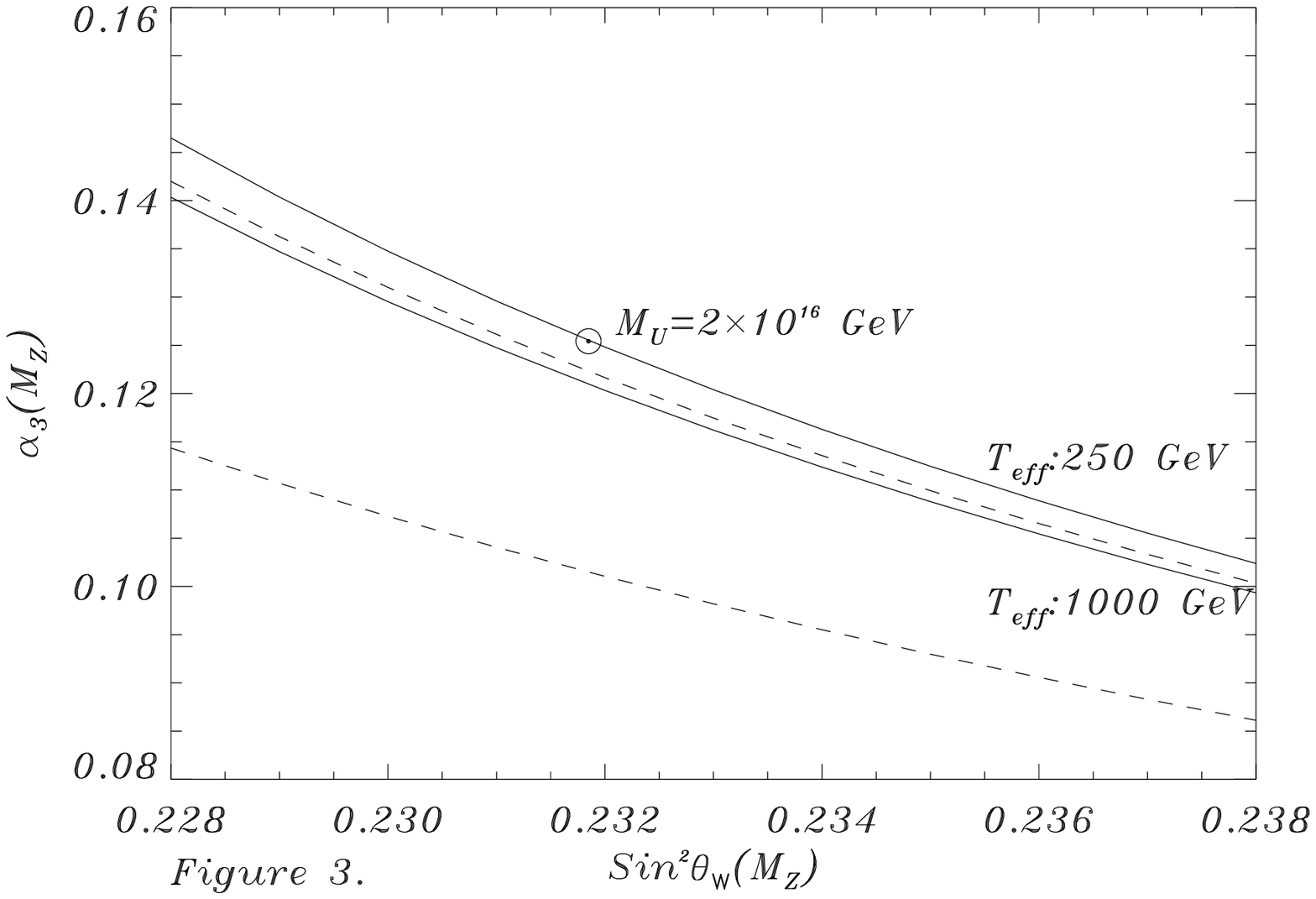,height=7.5cm,width=7.5cm}} 
\end{tabular} 
\newline 

{\small Figure 2. The values of the derivative ${d\Delta ^{H}}/{d(\ln \mu 
_{0})}=1-2x\ln _{x}^{\prime }\eta \left( i{3\sqrt{3}\pi }{e^{\gamma _{E}-1}}%
/(2x^{2})\right) $ and that of $\Delta ^{H}$ for $x$ close to 1. Figure 3. 
The values of $\alpha _{3}(M_{Z})$ for the model with gauge bosons in the 
N=2 sector with a change of $x$ from 1 to 1/2, leading to an area (between 
dashed lines) of values available to $\alpha _{3}(M_{Z})$ larger by a factor 
of $\approx 5$ compared to the MSSM case (continuous lines). } 
\end{figure} 
To understand the importance of $\Delta ^{H}$ and the implications it has 
for the threshold sensitivity of $\alpha _{3}(M_{Z})$, we note that $\Delta 
^{H}$ plays a central role in the attempts to bridge the well-known ``gap'' 
(of a factor of $\approx 20$) \cite{dienes,nilles1} between the heterotic 
string scale and the MSSM unification scale. It has been found (for a review 
see \cite{nilles1}) that for generic examples ($Z_{8}$ orbifold) this 
condition requires $\Delta \approx 8.5$ \cite{nilles1,nilles2} which in turn 
requires moduli of considerable size, $T_{2}\approx 18.7$ corresponding to a 
small value of $x\approx 0.53$ \footnote{%
This values for $T_{2}$ can still be on the edge of the weakly coupled 
regime of the 10D heterotic string as its coupling is equal to $\lambda 
=(\prod_{i}ReT_{i}/ReS)^{1/2}$)}. As may be seen from Figure 2 this leads to 
an even larger value for $|d\Delta ^{H}/d(\ln x)|$ and an enhanced relative 
threshold sensitivity. For the $Z_{8}$ orbifold \cite{ibanez46}, we find 
(using ${\overline{b}_{a}}=(-9/2,-5/2,-3/2)$) ${\cal {R}}=5.9$ for $x\approx 
0.53$. Since ${\cal {R}}$ gives the relative sensitivity of the gauge 
coupling predictions to the N=2 threshold and the SUSY threshold we see that 
even for $x=0.5$ the precision of the gauge coupling prediction is 
substantially reduced. The requirement that this precision should not be 
lost constrains $x\approx 1$ (i.e.$\mu _{0}=M_{string}$). However this fails 
to ``bridge the gap''. 
 
This example clearly illustrates the general problem that weakly coupled 
heterotic string models have to reconcile two conflicting constraints, 
namely the need for a large $\Delta ^{H}$ to solve the scale mismatch 
between the MSSM unification scale and the heterotic string scale and the 
need for a small derivative of $\Delta ^{H}$ to avoid a large threshold 
sensitivity of the gauge couplings. The latter constraint is introduced by 
the power-law dependence\footnote{%
This is due almost entirely to the presence of the towers of Kaluza-Klein 
modes rather than to winding modes, \cite{string}.} of the thresholds on the 
compactification scale, and even though such running takes place over a 
small energy range it still gives significant effects. The problem can be 
avoided if $x\approx 1$, corresponding to the \ compactification scale being 
very close to the string scale. On the other hand the former constraint 
seems to \ require a small value for $x.$ As we shall discuss this 
conclusion may be evaded in theories with Wilson line breaking. 
 
These conclusions apply to string theories with an $N=2$ sector. It is 
possible to construct string theories in which this sector is absent in 
which case there is no significant one-loop sensitivity of the unification 
prediction for the gauge couplings to changes of the high scale/moduli 
fields. The downside is that then one does not have the large $\Delta ^{H}$ 
needed to ``close the gap''. Even if this problem is solved in another way 
the $N=4$ sector necessarily present introduces strong threshold dependence 
at two loop order which provides almost as restrictive a bound on the 
compactification scale. We shall discuss this further in Section \ref 
{twoloop}. 
 
\subsection{Weakly coupled heterotic models with Wilson lines.} 
 
\label{wilson} In the case that there are Wilson lines characterised by the 
moduli $B$ the masses of the heavy states contributing to the RG flow of the 
gauge couplings become $B$-dependent. For this class of models the string 
calculation of the threshold contribution to the gauge couplings was 
computed in \cite{stieberger}. The presence of the Wilson lines may or may 
not break the symmetry group. In the latter case the threshold corrections 
do not seem to be large enough \cite{nilles2, nilles1} to account for the 
energy gap between the MSSM unification scale and that of the weakly coupled 
heterotic string. The case when Wilson lines break the symmetry gauge group 
is more interesting, as one can obtain a symmetry group close to that of the 
standard model \cite{standardmodelgroup}. We will consider only the class of 
models with (0,2) compactifications which have the additional advantage of 
avoiding the doublet-triplet splitting problem. For the case of the $Z_{8}$ 
orbifold the threshold contribution is then \cite{stieberger} (see also \cite 
{nilles1})  
\begin{equation} 
\Delta ^{H}=-\frac{1}{20}\ln \left[ Y^{10}\left| \frac{1}{128}%
\prod_{k=1}^{10}\theta _{k}(\Omega )\right| ^{4}\right]   \label{wilsonline} 
\end{equation} 
with $\theta _{k}$ the ten even, genus two theta functions. As has been 
shown in \cite{nilles1,stephan,nilles2}, in this case one can 
obtain a large value 
of the threshold $\Delta ^{H}$ which may help solve the mismatch between the 
MSSM unification scale and that of the heterotic string theory, without the 
need of large moduli. For example one obtains\footnote{%
with our normalisation for $\Delta ^{I}$}, $\Delta ^{H}\approx 8.5$ for $%
T_{2}=U_{2}=4.5$ and $B=1/2$. The derivative of $\Delta ^{H}$ with respect 
to $T_{2}$ can easily be taken by using the expansion of (\ref{wilsonline})  
\cite{stieberger} up to ${\cal {O}}(B^{4})$, to give  
\begin{eqnarray} 
\frac{\delta \Delta ^{H}}{\delta \ln T_{2}} &=&-\frac{1}{2}\left\{ \frac{%
4U_{2}T_{2}}{4U_{2}T_{2}-B^{2}}+\frac{24}{5}\frac{d\ln |\eta (iT_{2})|}{d\ln 
T_{2}}-\frac{12}{5}\frac{B^{2}T_{2}d_{T_{2}}^{2}\ln \eta 
^{2}(iT_{2})d_{U_{2}}\ln \eta ^{2}(iU_{2})}{1-6B^{2}d_{T_{2}}\ln \eta 
^{2}(iT_{2})d_{U_{2}}\ln \eta ^{2}(iU_{2})}\right\}  \\ 
&\approx &-\frac{1}{2}\left\{ 1+\frac{24}{5}\frac{d\ln |\eta (iT_{2})|}{d\ln 
T_{2}}\right\}  
\end{eqnarray} 
where $d_{T_{2}}$ stands for a derivative with respect to $T_{2}$ and where 
the approximation used above holds for the particular point in the moduli 
space giving the right size of the threshold $\Delta ^{H}$. One may observe 
that this expression is very close to that of (\ref{hetderiv}), with the 
difference that one has smaller moduli than in the case without Wilson lines 
present. For the $Z_{8}$ orbifold at the point in moduli space given above 
we obtain ${\cal {R}}=1.50$ which is lower than in the case without Wilson 
lines by a factor of $\approx 4$ for the same value of the threshold $\Delta 
^{H}$. 
 
The overall conclusion one may draw from this example is that the Wilson 
line background can give a threshold $\Delta H$ which is large enough to 
obtain the correct unification scale with the additional benefit that its 
derivative with respect to $T_{2}$ is not significantly affected, leading to 
a significantly lower threshold sensitivity than in the other cases without 
Wilson lines. Such models can preserve the precision prediction for gauge 
coupling and give good agreement between the string unification scale and 
the MSSM value. 
 
\subsection{Type I/I$^{\prime }$ models with a N=2 sector.} 
 
In this case the corrections to the gauge couplings come from the N=1 
(massless) sector and from the N=2 massive winding (Type I$^{\prime }$) 
sector. The massive {N}=2 threshold corrections to $\alpha _{i}^{-1}$ at the 
string scale $M_{I}$ were computed in \cite{dudas} giving  
\begin{equation} 
\Delta ^{I}=-\frac{1}{2}\sum_{k}{\overline{b}_{i,k}}\ln \left( \sqrt{G_{k}}%
ImU_{k}M_{I}^{2}|\eta (U_{k})|^{4}\right)  \label{typeI} 
\end{equation} 
Here $G_{k}$ is the metric on the torus\footnote{%
We consider in the following only compactification on a single torus, $T^{1}$%
.} $T^{k}$, and $\sqrt{G_{1}}={R}_{1}{R}_{2}$ (for a rectangular torus). The 
behaviour of the thresholds when $ImU={R}_{1}/{R}_{2}$ is of order one is 
logarithmic, $\Delta _{a}\sim \ln ({R}_{1}{R}_{2})$ and when $ImU={R}_{1}/{R}%
_{2}\gg 1$ it is power-like, $\Delta _{a}\sim R_{1}/R_{2}$. 
 
The interest in this class of models was initially due to the observation 
that, in contrast to the heterotic case, the $\Delta ^{I}$'s have 
logarithmic running which, even for a low string scale, may continue up to 
the Planck scale \cite{bachas} (more precisely up to the first winding mode 
close to this scale). Despite the absence of power-like terms in (\ref{typeI}%
), such models still have fine-tuning problems and this has been extensively 
discussed in \cite{gauge_unif}. The origin of the fine tuning is the 
difficulty in keeping light the closed string state with vacuum quantum 
numbers (dual to the first winding mode) which provides the cut-off for the 
gauge coupling running. Its mass is $M_{I}^{2}/M_{P}$ and to obtain running 
to the MSSM gauge unification scale $M_{I}$ should be large ($>10^{10}$ GeV). 
 
In the following we will determine the relative precision factor ${\cal {R}}$%
, induced by the structure of the thresholds (\ref{typeI}), assuming that 
the linear combination of beta functions entering the definition of ${\cal R} 
$ is of order ${\cal O}(1)$. The case this is not true is discussed in 
Section \ref{twoloop}. The variation of the thresholds with respect to $%
r_{1,2}=M_{I}R_{1,2}$ where $R_{1,2}$ are the compactification scale(s) 
gives the relative fine tuning measure  
\begin{equation} 
{\cal R\simeq }\frac{d\Delta ^{I}}{d(\ln r_{1})}=-\frac{1}{2}-2\rho \frac{d}{%
d\rho }\ln \eta (i\rho )  \label{typeIR} 
\end{equation} 
where $\rho =r_{1}/r_{2}$. Quantitatively ${\cal R}$ $\approx 10$ for $%
r_{1}/r_{2}<1/20$ or $r_{1}/r_{2}>20.$ Since this significantly degrades the 
precision we are led to the conclusion that anisotropic compactifications 
are severely limited for reasonable values of the beta coefficients. Models 
with large radii of compactification are apparently still allowed as it is 
possible to have $r_{1}$ and $r_{2}$ very large while keeping $\rho \cong 
r_{1}/r_{2}\approx {\cal O}(1)$. However they suffer from an even more 
severe problem in that the $N=2$ beta functions have to be proportional to 
the $N=1$ beta functions if the MSSM\ predictions are to be preserved\cite 
{gauge_unif}. We should stress that this sector does {\it not} correspond to 
Kaluza Klein excitations of the Standard Model states and so to obtain 
negative beta functions it is necessary that it involves a new massive gauge 
sector which also carries Standard Model gauge quantum numbers. Even with 
this it is very difficult to construct a spectrum giving beta functions 
proportional to the Standard Model beta functions because a N=2 copy of the 
MSSM representations gives a contribution to the beta functions which is  
{\it not} proportional to the $N=1$ contribution \cite{gauge_unif}. Even 
more implausible is the possibility that two-loop corrections should be the 
same as in the MSSM. Thus in this case the good prediction of the MSSM must 
be viewed as a complete accident as the running above the string scale comes 
from a spectrum completely unrelated to that of the MSSM. 
 
\subsection{Type I/I$^{\prime }$ models without a N=2 sector} 
 
For models of this class we address a generic type I string inspired model  
\cite{aldazabal} which shares some similarities with the MSSM because it 
only has logarithmic running, but the spectrum of the light states and the 
hypercharge normalisation are different. Somewhat unexpectedly, we will show 
that logarithmic RG evolution is not always a sufficient condition for a low 
threshold sensitivity of $\alpha_3(M_Z)$ prediction. This is so because the 
definition of ${\cal {R}}$ depends not only on the scale, but also on the 
beta function coefficients which in the model to discuss will play a more 
important role following the non-standard hypercharge normalisation. 
 
The model is derived from D=4, N=1 compact type IIB orientifolds with $D_{p}$ 
branes and anti-$D_{p}$ branes located at different points of the underlying 
orbifold \cite{aldazabal}. It has gravity mediated supersymmetry breaking, 
full Standard Model gauge group, has no N=2 sector and the UV cut-off scale 
for the N=1 sector is the string scale. The predictions of the model for the 
gauge couplings unification were discussed at one loop order in \cite 
{aldazabal} and at two loop level in\cite{gauge_unif}. It involves 
logarithmic unification at a low value of the string scale ($\approx 10^{12}$ 
GeV). Here we address the issue of the relative sensitivity ${\cal R}$ of 
the prediction for $\alpha _{3}(M_{Z})$ (or $\sin ^{2}\theta _{W})$ to the 
heavy thresholds compared to the SUSY thresholds. The reason a low 
unification scale is possible in such models is because the (string) 
unification relation between gauge couplings is not that given by $SU(5)$ 
due to a different hypercharge normalisation$.$ To compensate for this it is 
necessary to extend the MSSM light spectrum to include five further pairs of 
Higgs doublets (of mass $\tilde{m}$) and three vector-like right handed 
colour triplets $(d+d^{c})$. When the latter have a mass equal to $\tilde{m}$ 
as well, the full two loop RGE equations derived from (\ref{rge_general}) 
are given by \cite{gauge_unif}  
\begin{equation} 
\alpha _{i}^{-1}(M_{Z})=-\delta _{i}+\alpha _{g}^{-1}+\frac{b_{i}}{2\pi }\ln %
\left[ \frac{M_{s}}{M_{Z}}\right] +\frac{\delta b_{i}}{2\pi }\ln \left[  
\frac{M_{s}}{\tilde{m}}\right] +\frac{1}{4\pi }\sum_{j=1}^{3}Y_{ij}\ln \left[ 
\frac{\alpha _{g}}{\alpha _{j}(\tilde{m})}\right] +\frac{1}{4\pi }%
\sum_{j=1}^{3}\frac{b_{ij}}{b_{j}}\ln \left[ \frac{\alpha _{g}}{\alpha 
_{j}(M_{Z})}\right]  \label{HMSSM} 
\end{equation} 
where $Y_{ij}={\delta b_{j}}/({b_{j}b_{j}^{\prime }})(2b_{j}T_{j}(G)\delta 
_{ij}-b_{ij})$, with $T_{j}(G)=\{0,2,3\}_{j}$; $b_{ij}$ is the two loop beta 
function as in the MSSM but with $3/11$ hypercharge normalisation, $%
b_{j}=\{11\xi ,1,-3\}_{j}$, ($\xi =3/11$), $b_{j}^{\prime }=b_{j}+\delta 
b_{j}$, with $\delta b_{j}=\{7\xi ,5,3\}$ to account for the additional five 
Higgs pairs and three pairs of right handed colour triplets. 
 
The value of ${\cal {R}}$ is easily computed at one loop giving  
\begin{equation} 
{\cal {R}}=\frac{14}{19}\left( 2\delta b_{1}-3\delta b_{2}+\delta 
b_{3}\right) \approx 6  \label{ratiofine} 
\end{equation} 
where the definition of ${\cal {R}}$, (similar to that of (\ref{sensitiv}) 
with $\mu _{0}=\tilde{m}$, $M_{s}$ fixed) takes into account the different 
hypercharge normalisation of the model. This value for ${\cal {R}}$ implies 
a threshold sensitivity with respect to $M_{s}/{\tilde{m}}$ significantly 
greater than that of the MSSM with respect to $T_{eff}$. This may also be 
seen in the full two loop analysis which is used in Figure 4(a) and shows 
(at a global level) that a change of the ratio $M_{s}/\tilde{m}$ by a factor 
of 4, to match the uncertainty\footnote{corresponding
to a change of $T_{eff}$ between 250 to 1000 GeV.}
 in $T_{eff}$, gives a much larger variation 
than that found due to SUSY\ threshold uncertainty. The ratio of the two 
areas is $\approx 6.$  
\begin{figure}[t] 
\begin{tabular}{cc|cr|} 
\parbox{8cm}{ 
\psfig{figure=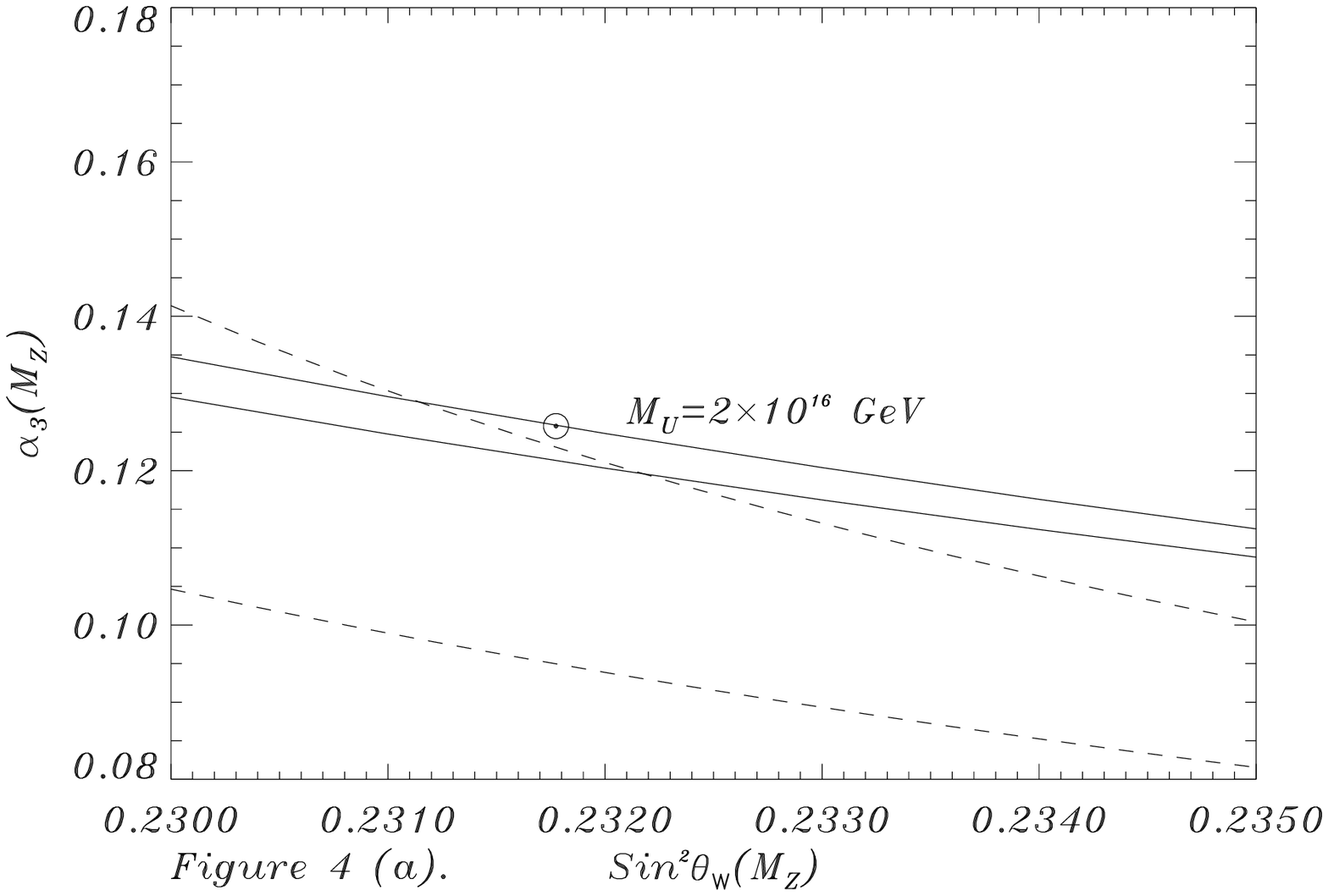,height=7.5cm,width=7.5cm}}  
\hfill{\,\,\,\,\,\,\,\,\,\,\,\,\,\,\,}  
\parbox{8cm}{ 
\psfig{figure=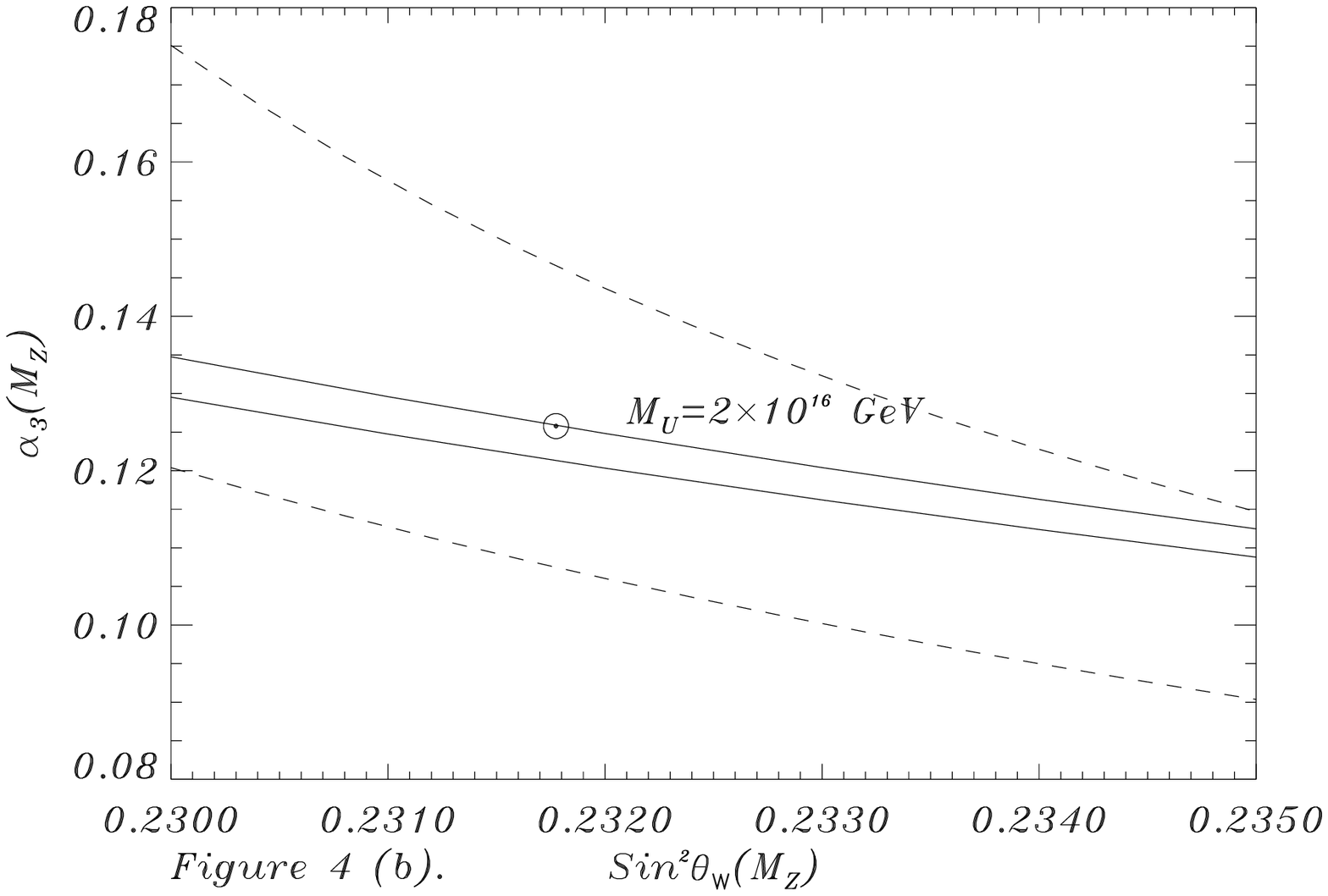,height=7.5cm,width=7.5cm}}  
\end{tabular} 
\newline 

{\small {Figure 4 (a): The values of $\alpha_3(M_Z)$ in function of $%
\sin^2\theta_W$. The widely separated curves (dashed line) correspond to the 
model investigated where the bare mass of the doublets and triplets 
(considered degenerate) is changed by a factor of 4. The area between these 
two curves is far larger than that of the MSSM (between the continuous 
lines) for a similar ratio (of 4) of the supersymmetric effective scales ($%
250<T_{eff}<1000$ GeV), meaning a less predictive power than in the MSSM 
case.\newline 
Figure 4 (b): The values of $\alpha_3(M_Z)$ in function of $\sin^2\theta_W$. 
The dashed lines correspond to the present model when the bare mass of the 
triplets $\approx 4 \tilde m$ meaning a less predictive power than in the 
previous case or than in the MSSM (area between the continuous curves) 
induced by an equal factor of change of the supersymmetric threshold from $%
250$ to $1000$ GeV. The slopes of the curves for the string model are 
different from those for the MSSM case due to the different hypercharge 
assignment.}} 
\end{figure} 
The problem is even worse if one relaxes the assumption of equal bare masses 
for the five Higgs pairs and colour triplets. 
 
This example illustrates the problems associated with any model which seeks 
to reduce the unification scale by going to a non-SU(5) normalisation for 
the weak hypercharge. While the increased threshold sensitivity and the 
associated loss of precision in the prediction for gauge couplings is bad 
enough we think an even more significant indictment of the scheme is that 
such models require that the precision agreement of the MSSM unification 
prediction with experiment is just a celestial joke! 
 
\section{Two loop limits on the compactification scale} 
 
\label{twoloop} So far we have discussed the one-loop threshold corrections 
coming from eq(\ref{rge_general}). We now turn to the case the linear 
combination  
\begin{equation} 
f(b_{i},{\overline{b}_{i}})=\overline{b}_{1}\frac{b_{23}}{b_{12}}+\overline{b%
}_{2}\frac{b_{31}}{b_{12}}+\overline{b}_{3} 
\end{equation} 
is zero and the one-loop corrections to the correlation between $\alpha _{3}$ 
and $\sin ^{2}\theta _{W}$ vanish. In such a case the heavy threshold 
correction only comes from two loops (and beyond). 
 
The two loop corrections come from the coefficients $Z_{\phi }(\Lambda ,Q)$ 
in eq(\ref{rge_general}). These are the one-loop matter wavefunction 
renormalisation coefficients above the scale $Q$. It is necessary to 
regulate the theory in order to determine the $Z_{\phi }(\Lambda ,Q).$ In 
the case of a GUT (effective) theory it is normally assumed that they are 
unity at the GUT scale (i.e. $\Lambda $ is chosen as the GUT scale). However 
this is not the correct prescription if the GUT emerges from string 
compactification for the string regularisation requires that $\Lambda $ be 
the string cut-off scale, usually the string scale. 
 
The coefficients $Z$ have two contributions. One is due to the usual (gauge 
and Yukawa) corrections from the N=1 ``MSSM-like'' massless states which 
generate the $Z_{\phi }(\Lambda ,Q)$ at one loop. These lead to corrections 
in the gauge couplings at ${\cal O}(\alpha ^{2}/(4\pi ))$. The second 
contribution to $Z_{\phi }(\Lambda ,Q)$ comes (for $Q>\mu _{0}$) from towers 
of (N=2,4) Kaluza Klein states. Such corrections of order ${\cal O}(N\alpha 
^{2}/4\pi )$ and will give ${\cal {R\approx }}$ $N\alpha /4\pi $. The 
constraint that ${\cal {R}}$ should not be large requires $N<4\pi /\alpha $ 
(equivalent to the requirement that the theory remain perturbative). While 
less constraining than the one-loop constraint, this is still very strong 
due to the rapid increase in the number of Kaluza Klein states, $N\approx 
(\Lambda /\mu _{0})^{\delta }$, $\delta $ being the number of extra 
dimensions. For $\delta =2,$ $\Lambda <10\mu _{0}$ while for $\delta =6,$ $%
\Lambda <2.5\mu _{0}.$ Thus, even if one can avoid the large threshold 
sensitivity (``power-like'' ) induced at one loop level by the (heterotic) 
string thresholds $\Delta $ (or at the effective field theory level by 
towers of Kaluza Klein states) through an (string) construction which sets $%
f(b_{i},{\overline{b}_{i}})=0$, the constraint on two-loop contributions 
still requires the cut-off (string scale) $\Lambda $ and the 
compactification scale $\mu _{0}$ should be very close. In particular string 
models with a Grand Unified group unbroken below the compactification scale 
should still have the compactification scale close to the string scale.

\section{Summary : A profile of a String Model\label{profile}} 
 
In this paper we have considered two issues. The first is the precision of 
the prediction in the MSSM for low-energy gauge couplings assuming an 
underlying unification at a high scale. Taking account of the uncertainties 
due to the unknown masses of the SUSY partners of Standard Model states we 
found that the predicted range for the gauge couplings compared to the a 
priori range of possible couplings is remarkably precise, between 2 and 
0.2\%. This gives a quantitative estimate of how significant is the gauge 
unification prediction. The prediction for $\sin^2 \theta _{W}$ itself is also 
impressive with an accuracy of $1.3 \%$. 
 
This remarkable precision led us to consider the nature of an underlying 
(string) theory that can maintain this accuracy. Due to the non-decoupling 
of the contribution of massive states to renormalisable terms in the low 
energy effective Lagrangian, the requirement that the gauge predictions be 
undisturbed places strong constraints on the massive sector. We determined 
the contribution of states transforming non-trivially under the Standard 
Model gauge group in compactified string models. Requiring that these 
contributions leave the MSSM predictions intact leads to a constraint on the 
magnitude of the compactification radius associated with the propagation of 
such states to limit the number of states in the Kaluza Klein tower. We 
found that for a wide class of string theories the contribution of heavy 
states occurs at one-loop order and the radius should be very close to the 
inverse of the string scale (well within a factor of 2). In the cases the 
one-loop contributions vanishes the limit is only slightly relaxed. 
 
The implication of this result is quite far-reaching. One immediate one is 
that unification at a low scale through power law running of the gauge 
couplings cannot maintain the precision of the MSSM predictions. Indeed, due 
to the different contributions to the beta functions of massive compared to 
massless SUSY representations, low scale unification requires a very 
different multiplet content from the MSSM in order to obtain the same gauge 
unification prediction. As a result the precision prediction for gauge 
couplings must be considered a fortuitous accident, something we find hard 
to accept given the remarkable precision of the prediction. Even if we do 
accept this, and the N=2 sector happens to give the same beta function as 
the N=1 MSSM spectrum, power law running introduces a strong dependence on 
the heavy thresholds so that the precision of the ``prediction'' is lost. 
For both these reasons we consider a low scale of unification due to power 
law running to be unlikely. Models with a low scale of unification due to a 
non-$SU(5)$ hypercharge normalisation do not require power law running. 
Nonetheless it turns out that they too have enhanced threshold dependence 
due to the need for additional light states and again this loses the 
predictive power of the MSSM. 
 
The profile of our string model which preserves the precision prediction for 
gauge couplings found in the MSSM therefore requires a large scale of 
unification with an $SU(5)$ normalisation for the weak hypercharge. Even so, 
there is still a strong constraint on the compactification scale because the 
cutoff of the contribution of heavy states is at the string scale or higher. 
To avoid the same power law running corrections that degraded the 
predictions in the case of low scale unification, the radius of 
compactification of those dimensions in which Standard Model gauge 
non-singlet fields propagate must not be large compared to the cutoff 
radius. This means the compactified string theory lies far from the 
Calabi-Yau limit and close to the superconformal limit. This is quite 
attractive in that many aspects of the effective low-energy field theory, 
such as Yukawa couplings, are amenable to calculation in the superconformal 
limit. 
 
The need for a high scale of unification broadly fits the expectation in the 
(weakly coupled) heterotic string. However in detail there is a discrepancy 
between the MSSM value for the unification scale of $3.10^{16}GeV$ and the 
prediction in the weakly coupled heterotic string, approximately a factor of 
20 larger. Three explanations have been suggested. 
 
The first is that heavy threshold effects raises the unification scale to 
the predicted value. The requirement that the precision of the gauge 
coupling prediction be maintained severely limits the magnitude of these 
threshold effects and precludes explanations requiring large radii. However 
in models with Wilson line breaking it is possible to have very large 
threshold corrections to the unification scale while keeping the threshold 
contribution to gauge couplings small. Given that such Wilson line breaking 
is very often needed to break the underlying gauge symmetry of the heterotic 
string, this explanation seems very reasonable. 
 
A second possibility is to have a GUT below the string compactification 
scale, the GUT breaking scale being the unification scale . Even in this 
case it is still necessary for the compactification scale to be very close 
to the string scale to keep the two loop contributions from the Kaluza Klein 
states small. If this theory comes from the weakly coupled heterotic string 
the compactification and string scales must be close to the Planck scale. 
 
Of course our analysis does not preclude the existence of large new 
dimensions not associated with the propagation of Standard Model states. A 
particular example is the strongly coupled heterotic string in which gravity 
but not the Standard Model states propagate in the eleventh dimension. For 
the case the eleventh dimension is three orders of magnitude larger than the 
string length, corresponding to a compactification radius of $O(10^{-14}fm),$ 
the gauge unification scale may be reduced to that found in the MSSM. This 
provides a third way to reconcile the predicted unification scale with that 
needed in the MSSM. However the size of the extra dimensions is still 
severely constrained by the need to have a high unification scale, the 
minimum occurring for just one additional dimension as in the strongly 
coupled heterotic case. 
Thus even in the case of large new dimensions in which 
Standard Model states do {\it not}
propagate, the new dimension cannot be larger than $10^{-14}$ fm.

\vspace{0.5cm} 
\noindent {\bf Acknowledgement} The work of D.G. was supported 
by the University of Bonn under the
European Commission RTN programme HPRN-CT-2000-00131 and
by  the University of Oxford (Leverhulme Trust research grant).
This research is also 
supported in part by the EEC research network HPRN-CT-2000-00148.

\end{document}